\documentclass[superscriptaddress,twocolumn,10pt,nofootinbib,floatfix]{revtex4-1}
\usepackage{amssymb}
\usepackage{amsmath}
\usepackage{appendix}
\usepackage{times}
\usepackage{graphicx}
\usepackage{subeqnarray}
\usepackage{bbold}
\usepackage{enumerate}

\usepackage{color}
\usepackage{bm}

\begin{document}
\title{Ferroelectric higher-order topological insulator in two dimensions}

\author{Ning Mao}
\author{Runhan Li}
\author{Xiaorong Zou}
\author{Ying Dai}
\email{daiy60@sdu.edu.cn}
\author{Baibiao Huang}
\affiliation
{School of Physics, State Key Laboratory of Crystal Materials, Shandong University, Jinan 250100, China}
\author{Chengwang Niu}
\email{c.niu@sdu.edu.cn}
\affiliation
{School of Physics, State Key Laboratory of Crystal Materials, Shandong University, Jinan 250100, China}

\begin{abstract}
The interplay between ferroelectricity and band topology can give rise to a wide range of both fundamental and applied research. Here, we map out the emergence of nontrivial corner states in two-dimensional ferroelectrics, and remarkably demonstrate that ferroelectricity and corner states are coupled together by crystallographic symmetry to realize the electric control of higher-order topology. Implemented by density functional theory, we identify a series of experimentally synthesized two-dimensional ferroelectrics, such as In$_2$Se$_3$, BN bilayers, and SnS, as realistic material candidates for the proposed ferroelectric higher-order topological insulators. Our work not only sheds new light on traditional ferroelectric materials but also opens an avenue to bridge the higher-order topology and ferroelectricity that provides a nonvolatile handle to manipulate the topology in next-generation electronic devices.
\end{abstract}

\maketitle
\date{\today}

Time-reversal polarization, defined as the differences between Wannier charge centers of Kramers pair, plays a pivotal role in the $\mathbb{Z}_2$ classification of electronic topological insulators (TIs)~\cite{Qi2011,Hasan2010,Bansil2016rmp}. Armed with topologically protected metallic boundary (surface or edge) states, TIs ensure the dissipationless nature of spin transport with promising applications in spintronics~\cite{Kane,bhz,z2fuliang,she}. The existence of boundary-localized mass domains, as generally accepted, will open up a boundary band gap, thereby trivializing the time-reversal polarization in TIs. Remarkably, recent investigations proposed that the hinge or corner states may appear, resulting in the intriguing quantum phase named higher-order TIs (HOTIs)~\cite{Benalcazar61,benalcazar2017electric,highorderfirst,maonpj,maoas}. For which, $nth$-order HOTIs in $d$ dimensions host symmetry protected features not at the ($d-1$)-dimensional but at ($d - n$)-dimensional boundaries. In particular, as the expectation value of position operator for Wannier functions, the quantized polarization can be topological indices for $C_n$-symmetric HOTIs, which form into $\mathbb{Z}_2$, $\mathbb{Z}_2 \times \mathbb{Z}_2$, or $\mathbb{Z}_3$ indexes~\cite{Song2018,po2017symmetry,Tang2019}. As a novel extension of TIs, theoretical models and material candidates of HOTIs have been proposed both in two/three dimensions~\cite{highorderezawa1,higherordereuln2as2, Yue2019,Park216803,highordergraprl,highordersplit,highordergranpj,highorderbieuo,highorderrenyafei}, and remarkably been experimentally observed in three-dimensional Bi~\cite{frankbi}, WTe$_2$~\cite{Choi2020NM}, and Bi$_4$Br$_4$~\cite{noguchi2021evidence}. However, for two-dimensional electronic materials, the experimental confirmation of HOTIs has been elusive so far.

On the other hand, polarization has received great attention in ferroelectric materials~\cite{ ferro, ferro_review1,ferro_review2, ferrobeginner,FTI1,FTI2,FTI3,FE1,FE2,FE3,FE4,FE5}. One important characteristic of the ferroelectric materials is the existence of two degenerate structures with opposite polarization that could be switched by an external electric field~\cite{ferro_review3}. In fact, the switchable electric polarization, when coupled to other properties such as ferromagnetism, ferroelasticity, valley, skyrmion, spin polarization, and/or nontrivial band topology, provides a nonvolatile control with the implementation of gate voltages~\cite{ferro_review4,ferro_mag, ferro_elas, vs2_bilayer,ferro_sky,ferro_spin,ferro_topo,ferro_yanliang}. Moreover, the separation of the center of positive and negative electric charge in ferroelectric systems is currently served as a hallmark of obstructed atomic limits (OALs) within the framework of topological quantum chemistry theory~\cite{ tqc1,tqc3}. The OALs have recently attracted significant interest, since they could manifest themselves at the boundary in terms of metallic surface/edge states, regardless of spin-orbit coupling (SOC) or bulk band inversion~\cite{xu2021three,xu2021filling,Nelson2021PRL,schindler2021non,unconventional,oaicatalytic}. Even more, some of them could host hinge/corner states as a result of filling anomaly, which leads to the higher-order topology. Therefore, a natural question arises as to whether the ferroelectric polarization can result in the emergence of HOTIs, especially in two dimensions.

In the present work, we demonstrate the emergence of higher-order topological phases in 2D ferroelectrics including both the in-plane and out-of-plane ferroelectricity. Moreover, we reveal that the in-plane polarizations, which can enforce the non-quantized ones and the quantized ones, endow the emergence of the corner states and even may serve as the topological invariants. Effective models for both the quantized and non-quantized in-plane polarizations are constructed to demonstrate the feasibility of attaining the proposed 2D ferroelectric HOTIs. Remarkably, based on the first-principles calculations, a wide range of 2D ferroelectrics are testified to possess fractional corner charges, resulting in higher-order topological phases. For which, the implementing of an external electric field could switch the directions of ferroelectric polarization and in turn change the position of corner states and/or mediates the topological phase transitions. Our results indicate that 2D ferroelectric materials provide an up-and-coming platform to achieve and control higher-order corner states with experimentally feasible examples.

It has long been known that polarization is closely related to the position of the Wannier function, which can be transformed into the Berry phase of the occupied bands as 
\begin{align}
	\textbf{P} = P^{occ} \hat{x} P^{occ} =-\frac{1}{2 \pi} \int_{k}^{k+2 \pi} \operatorname{Tr}\left[\mathcal{A}_{k}\right] d k,
\end{align}
where $P^{occ}$ and $\hat{x}$ are the projection and position operators, respectively. $\mathcal{A}_{k}  $ is the gauge-dependent Berry connection in terms of $\left[\mathcal{A}_{k}\right]^{m n}=-i\left\langle u_{k}^{m}\left|\partial_{k}\right| u_{k}^{n}\right\rangle$, and its integral along a closed path is called as Berry phase. The direction of polarization is determined by the displacement of Wannier function, 
and the polarization can be quantized or non-quantized limited by the crystal symmetry. Remarkably, as we demonstrate below, an in-plane polarization, no matter quantized or non-quantized, would give rise to the two-dimensional higher-order corner states.

\begin{figure}
	\centering
	\includegraphics{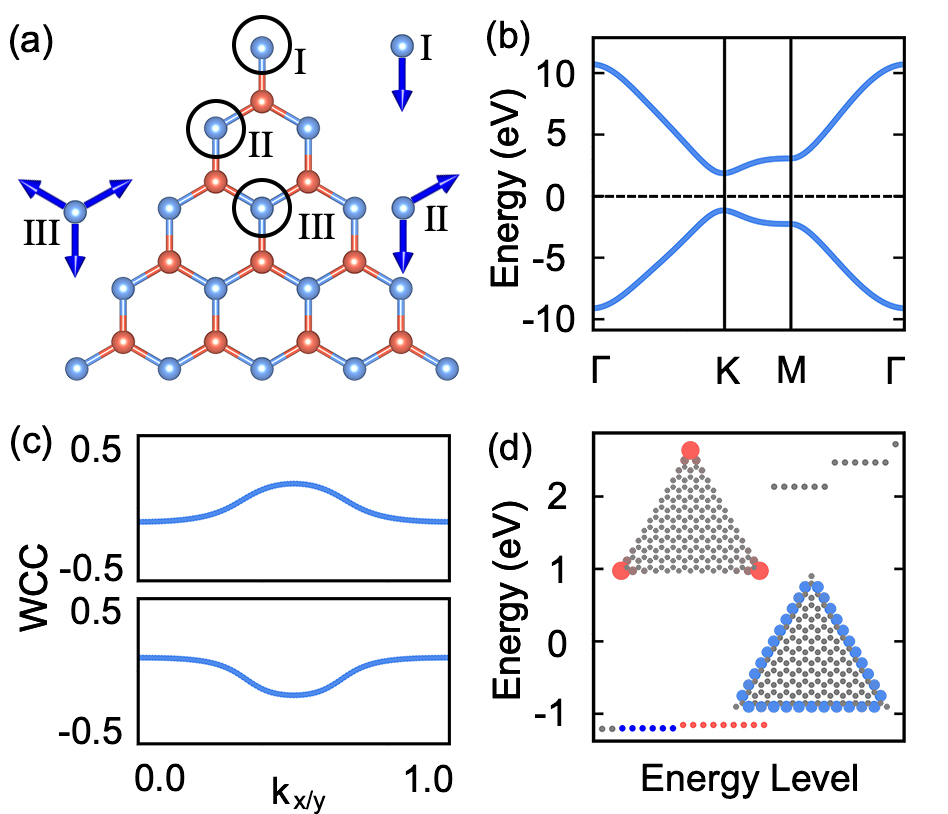}
	\caption{(a) Sketch of the finite triangular lattice with $C_3$ symmetry. The quantized polarization in the bulk (III) can always be annihilated, whereas the edge (II) and corner (I) may possess uncompensated polarization. (b) Band structures of the two-band Hamiltonian, and the Fermi level is indicated with a dashed line. The parameters are set as $\lambda_1$ = 2, $\lambda_2$ = -1, $t_1$ = -3, $t_2$ = 0.1, and $t_3$ = -0.2, respectively. (c) The Wilson bands of the occupied band along $k_x$ (up) and $k_y$ (down) directions. (d) Energy spectrum of the finite nanoflake. The red dots near the Fermi level represent the in-gap corner states, and the blue ones denote the edge states. The distributions of corner and edge states are plotted in the inset.}
	\label{1model}
\end{figure}

We consider first the quantized polarization in a two-band tight-binding model on a honeycomb lattice
\begin{align}
H= \sum_{i} \lambda_{i} c_{i}^{\dagger} c_{i} + t_1 \sum_{\langle i j\rangle} c_{i}^{\dagger} c_{j} + t_2 \sum_{\langle\langle i j\rangle\rangle} c_{i}^{\dagger} c_{j} + t_3 \sum_{\langle\langle\langle i j\rangle\rangle\rangle} c_{i}^{\dagger} c_{j}.
\end{align}  
The first term is the sublattice potential, which is needed for the broken of inversion symmetry ($\mathcal{I}$). The second, third, and fourth terms represent the nearest, next-nearest, and next-next-nearest neighbor hoppings, as depicted in Fig.~\ref{1model}(a). In Fig.~\ref{1model}(b), we show the band structures of the Hamiltonian under the OAL phase. Indeed, the moving of Wannier charge centers leads to a quantized polarization as shown by Fig.~\ref{1model}(c). However, the sum of the polarization, bounded by $C_3$ symmetry, can necessarily be zero in the bulk due to $\textbf{P}_{\rm bulk} = \sum_{i}( \textbf{P}_i + C_3 \textbf{P}_i + C_3^{-1} \textbf{P}_i )= 0$, while ones at the edges and corners cannot be offsetted as displayed in Fig.~\ref{1model}(a), leading to the non-vanished polarization as $\textbf{P}_{\rm edge} = \sum_{i}( \textbf{P}_i + C_3 \textbf{P}_i) $ and  $\textbf{P}_{\rm corner} = \sum_{i} \textbf{P}_i $. Hence electrons from the bulk will be driven to the edges and corners, forming the edge and corner states. Figure~\ref{1model}(d) presents the energy level of a finite triangular flake. Clearly, the edge and corner states emerge near the Fermi level, and the energy level of edge states is lower than corner states caused by the difference of polarization in the edge and corner.  Moreover, corner states originated from the valence and conduction bands are degenerated in energy, giving rise to a phenomenon of filling anomaly. In that sense, the corner states can be pushed into the conduction or valence bands as a whole, and the positive or negative charge will be generated to affect the charge neutrality.

For a better control of the corner states, the non-quantized polarization has to be introduced by breaking the $C_6, C_4,$ and $C_3$ symmetries. To study systematically the switchable corner states, we then focus on a simple model of one atom located at the center of a 2D square lattice~\cite{highorderezawa1}, and the Hamiltonian can be expressed as
\begin{align}
	\begin{split}
		H_0\!= [(m\!-\!t (\cos k_{x} \!+
		\!\cos k_{y})] \tau_{z} \!-\! \lambda( \sin k_{x}\sigma_{x} \!+\! \sin k_{y}\sigma_{y})\tau_{x}, 
	\end{split}
\end{align}
where $m$ sets the energy offset between $s$ and $p$ orbitals. Spin-orbit coupling (SOC) termed with $\lambda$ induces their hybridization while the nearest-neighbor hopping ($t$) bridges the relations for $s-s$ or $p-p$ orbitals. After diagonalizing the Hamiltonian and considering the zero-energy conditions for four high symmetry points of~$\Gamma$, X, Y, and M, the nontrivial TIs are expected to emerge as $|m/t| < 2$ (See Fig. S1~\cite{sm}). The non-quantized polarization affects the electric potential of the system and causes a potential imbalance for the overall system. In this sense, we add an electric potential term to mimic the non-quantized polarization 
\begin{align}
	\begin{split}
		H_{E}\!=\!\sum_{i} c_{i }^{\dagger}\left[e F(x \cos \theta+y \sin \theta)   \right] c_{ i },\\
	\end{split}
\end{align}
where $c^{\dagger}_{i} = ( c^{\dagger}_{i\uparrow}, c^{\dagger}_{i\downarrow})$ are electron creation operators at the site $i$. Besides the elementary charge $e$, $H_E$ is expressed with the amplitude given by $F$ and orientation defined by $\theta$. It is clear that, under an in-plane electric potential along $\hat{y}$ direction, $\mathcal{I}$, $C_6, C_4, C_3,$ and $C_{2x}$ symmetries are broken, and remarkably,  as illustrated in Fig.~S1~\cite{sm}, one can find that the corner states appear, serving as a direct signal of HOTI. 

\begin{figure}
	\centering
	\includegraphics{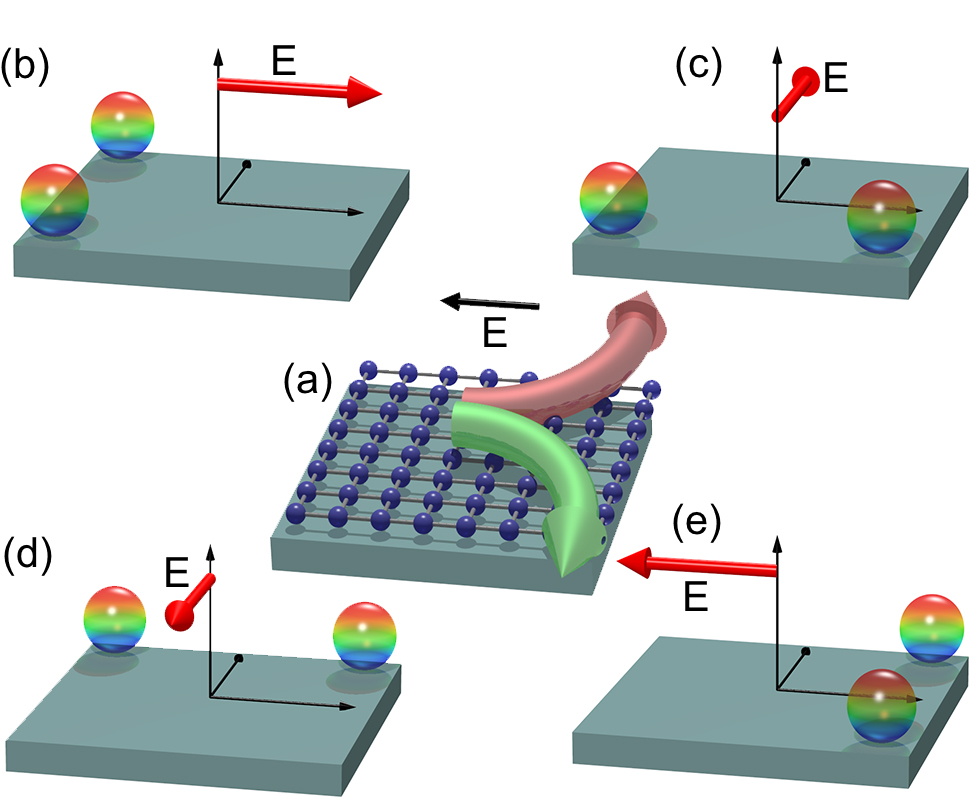}
	\caption{(a) In-plane polarization and (b)-(e), corner states under the in-plane electric field. A transverse current is generated under in-plane electric field, which leads to the in-plane polarization and then the birth of corner states in symmetric corners. Corner states are indicated with rainbow balls and arrows denote the directions of the in-plane electric field.}
	\label{model}
\end{figure}

However, the in-gap corner states appear only at specific corners, inspiring us to understand their physical origin. It is well known that electrons will acquire a transverse velocity in the presence of the in-plane electric fields, as schematically shown in Fig.~\ref{model}(a). This may lead to corner charge accumulation and thus manifest the corner states protected by the rotation symmetry. To map out the role of rotation symmetry and polarization, we choose four representatives with different in-plane orientations, where the rotation symmetries are highly sensitive to the orientations. In the absence of electric potential, the considered square lattice has both $C_{2x} = -i\sigma_{x}$ and $C_{2y} = -i\sigma_{y}$, but no polarization and then no corner states [see Fig.~S1~\cite{sm}]. If an electric potential is perpendicular to the $\hat{x} (\hat{y})$-direction, $C_{2y} (C_{2x})$ will be broken, and otherwise it will be conserved. For example in Fig.~\ref{model}(b), $C_{2y}$ is broken, while $C_{2x}$ survives and ensures the existence of corner states. Interestingly, the in-gap corner states can be tuned into opposite corners when the in-plane electric field as well as the polarization is reversed as shown in Fig.~\ref{model}(e). Similar results with corner states localized on $C_{2y}$-symmetric positions are shown in Figs.~\ref{model}(c) and~\ref{model}(d). Therefore, one can realize the nonvolatile control of the corner states.

\vspace{0.3cm}

\noindent{\bf  Material candidates.}
Having demonstrated the interplay of HOTIs and polarization, we aim now at its realization in electronic materials, where remarkably the in-plane polarization $\mathbf{P}$ can be defined by 
\begin{align}
	\mathbf{P} = p_1\mathbf{a_1} +  p_2\mathbf{a_2}.
\end{align}  
Here, $\mathbf{a_1}$ and $\mathbf{a_2}$ are primitive lattice vectors, and the components $p_1$ and $p_2$ can be served as the topological indices that are equivalent to the quantized Berry phase~\cite{cannotdelete}. The 2D ferroelectric materials lack the $\mathcal{I}$ intrinsically and hence usually possess the electric polarization, revealing the great opportunities for HOTIs in 2D ferroelectric materials. In addition, the ferroelectricity nature has been demonstrated experimentally in abundant 2D layered materials, and indeed the ferroelectric switching is currently maturing into a significant burgeoning research, therefore, it would be of great significance to make a bridge between HOTIs and 2D ferroelectricity.

\begin{figure}
	\centering
	\includegraphics{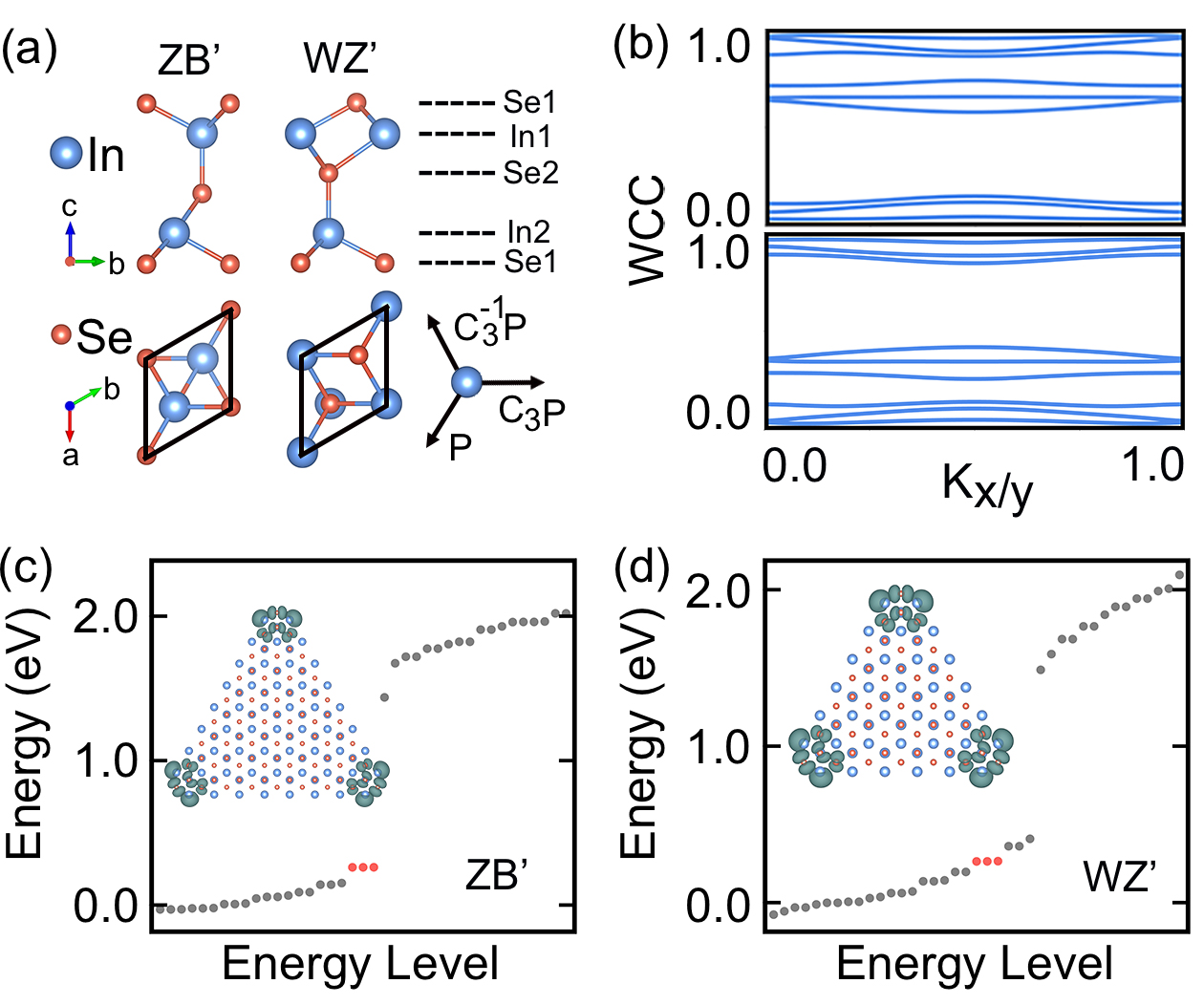}
	\caption{(a) Top and side views of In$_2$Se$_3$ in ZB' and WZ' phases. (b) Wilson bands of the occupied bands along $k_x$ (up) and $k_y$ (down) directions. Energy spectrums of In$_2$Se$_3$ in (c) ZB' and (d) WZ' phases. The red dots near the Fermi level represent the in-gap corner states. Distributions of corner states are displayed in the insets.}
	\label{in2se3}
\end{figure}

In$_2$Se$_3$ is one of the most famous examples of 2D ferroelectric materials, which has been widely explored with promising applications in phase-change memory, thermoelectric, photoelectric, and catalysis~\cite{in2se3_acs, in2se3NC, in2se3_nano1, in2se3_nano2,in2se3_cata}. The monolayer is stacking with a sequence of Se1-In1-Se2-In2-Se1 that can be easily realized by physical exfoliation and/or chemical vapor deposition. The ZB' and WZ' phases of In$_2$Se$_3$, as shown in Fig.~\ref{in2se3}(a), are the degenerate ground states that have already been synthesized~\cite{in2se3AM,in2se3NL1,in2se3NL2}. They belong to space group $P3m1$ (No.156). For ZB' phase, three Se atoms occupy two 1a and one 1c Wyckoff positions, whereas two In atoms occupy 1b and 1c Wyckoff positions. While three Se atoms occupy 1a, 1b, and 1c Wyckoff positions, and two In atoms occupy 1a and 1c Wyckoff positions for WZ' phase. 

According to the Wilson loop calculations as illustrated in Fig.~\ref{in2se3}(b), the electron transfer occurs when atoms are brought together to form the crystal of In$_2$Se$_3$, i.e., the electrons of In move to Se, giving rise to the in-plane electric polarization. Limited by $C_3$ symmetry, the polarization induced by the electron distribution between In1/In2 and Se2 atoms can be quantized as
\begin{align}
	\mathbf{P} = (1/3,1/3).
\end{align}
Although such a quantized polarization can be annihilated in the bulk as shown in Fig.~\ref{in2se3}(a), the uncompensated polarization remains intact at the corners, which renders the emergence of fractional charge with 2e/3 on the corners of $C_3$-symmetric finite lattice, and yields the coexistence of 2D ferroelectricity and higher-order topology, namely 2D ferroelectric HOTIs. To further identify the nature of HOTIs, we construct a triangular nanoflake of In$_2$Se$_3$, which preserves the $C_3$ symmetry for both the bulk and edge. As plotted in Figs.~\ref{in2se3}(c) and \ref{in2se3}(d), three degenerate in-gap states arise around the Fermi level, accumulating in three corners of Se2 atoms. However, the emergence of corner states between ZB' and WZ' phases are nearly the same, rendering a difficulty to control the corner states by an external electric field.

Aiming at revealing the nonvolatile control of corner states and the universality of pronounced ferroelectric HOTIs, we turn to the currently proposed sliding ferroelectricity, which remarkably reveals a universal approach and enables the design of 2D ferroelectric materials out of inherent polar compounds~\cite{ferro_review1,ferro_review2,ZrI2bilayer,BNbilayer_ex}. To date, such a theory has been successfully applied to a bunch of 2D binary compounds $XY$ and transition-metal dichalcogenides $XY_2$, such as BN, ZnO, AlN, GaN, SiC, InSe, GaSe, WTe$_2$, MoS$_2$, VS$_2$, and so on~\cite{bilayer_ferro}. In fact, all of these $XY$ and $XY_2$ are $C_3$-symmetric polar materials, where the electrons transfer from $X$ to $Y$. Therefore, a large quantized in-plane polarization is generated and drives the Coulomb interaction. If the bilayer lose mirror symmetry ($\mathcal{M}_z$) perpendicular to the z-axis, the Coulomb interaction will introduce a non-quantized out-of-plane polarization, making the systems to be the out-of-plane ferroelectric materials that can be controlled by the external electric field. 

\begin{figure}
	\centering
	\includegraphics{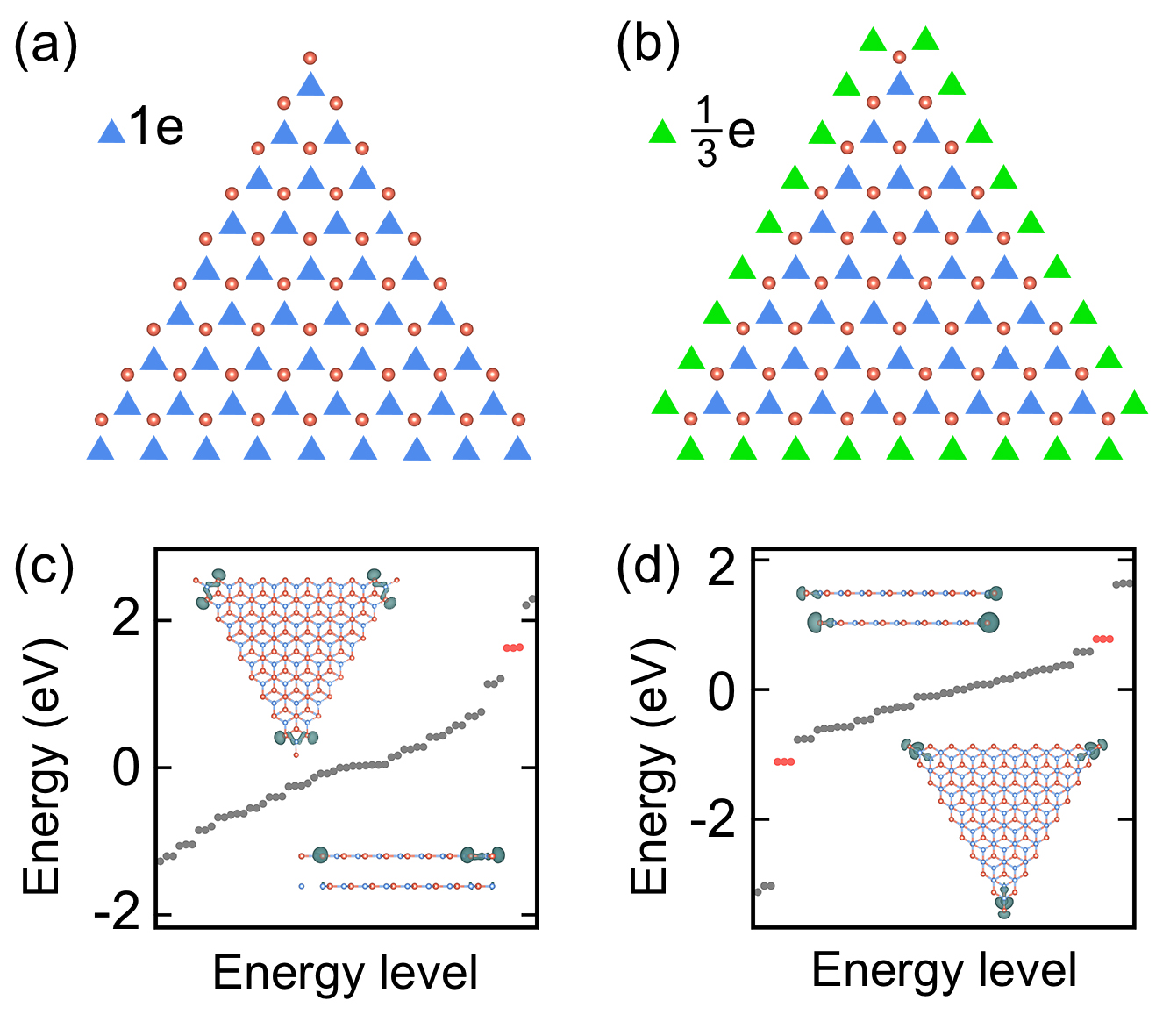}
	\caption{(a) Finite nanoflake with atoms at positions of red circles. Limited by the polarization, electrons of red circles move to positions of blue triangles, which is $C_3$-symmetric in the bulk but not in the corner. (b) Finite nanoflake formed by distributing the edge electrons shown in a, averagely in three edges. This configuration restores the $C_3$ symmetry in the corner, and leads to the 2e/3 corner charges. Energy spectrums of finite nanoflake for BN bilayer in (c) AB stacking and (d) BA stacking. Red dots represent the in-gap corner states, and the distributions of corner states are plotted in the insets.}
	\label{bn}
\end{figure}

We take boron nitride (BN) as an example. The valence atomic configurations for B and N atoms located at 1d and 1f Wyckoff position are $2s_22p_1$ and $2s_22p_3$, respectively, and when forming into the BN compound, all valence electrons of B are moved to the position of N, giving rise to the in-plane electric polarization along the B-N bond as
\begin{align}
	\mathbf{P} = (2/3, 1/3).
\end{align}
When considering only the polarization, the Wannier centers move from B atoms to the position of N atoms, leading to a neutral and integer charge configuration as schematically shown in Fig.~\ref{bn}(a). However, such a configuration is not $C_3$-symmetric at corners. To preserve the symmetry, all of the Wannier centers have to be equally distributed over the three sectors, leaving the edge charges in multiples of e/3 as illustrated in Fig.~\ref{bn}(b). Moreover, each corner is related to two fractional Wannier centers and thus manifests the corner charges of 2e/3. 

\begin{figure}
	\centering
	\includegraphics{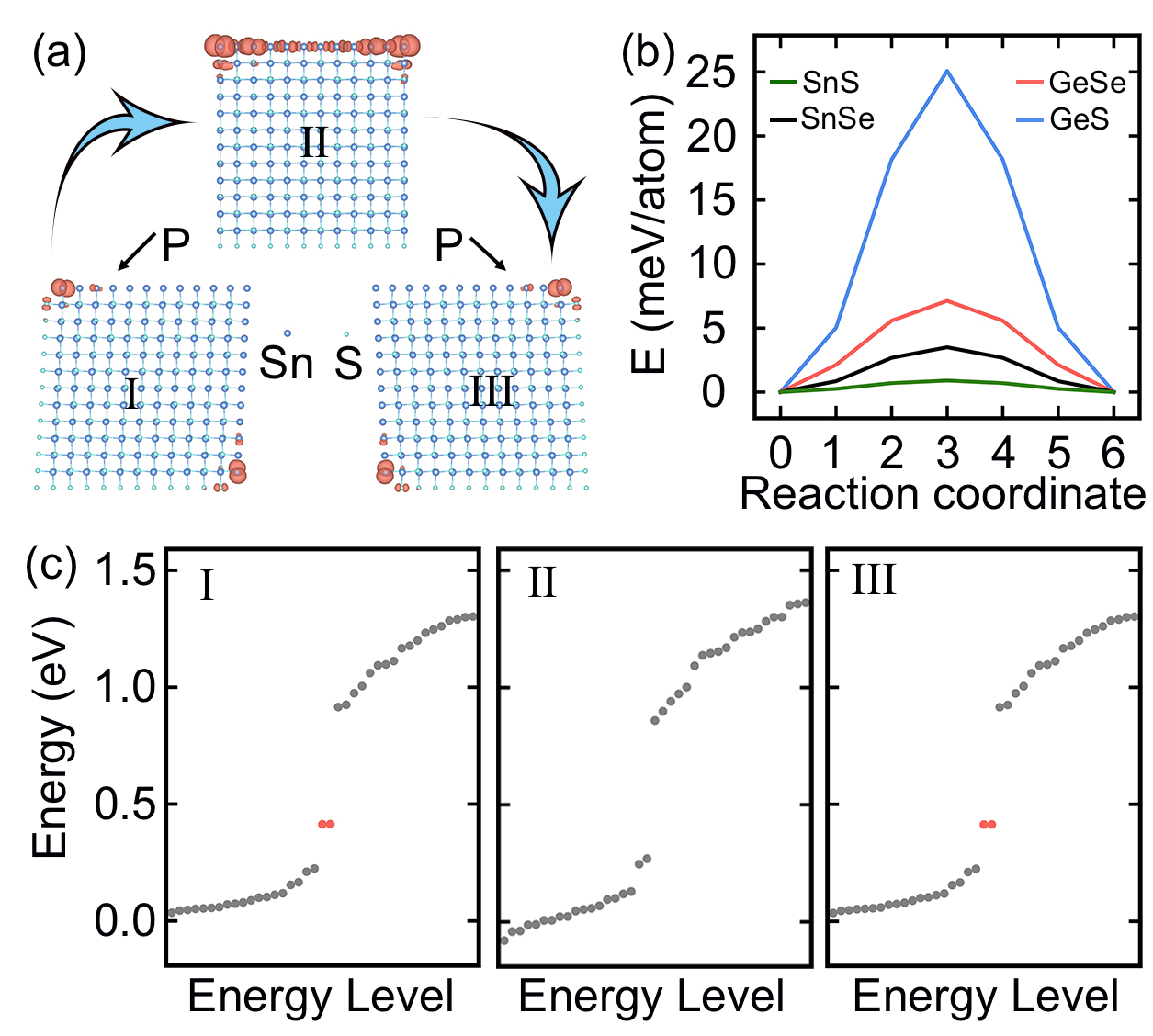}
	\caption{(a) Distributions of corner states for ferroelastic states \uppercase\expandafter{\romannumeral1}, \uppercase\expandafter{\romannumeral3}, and intermediate state \uppercase\expandafter{\romannumeral2} of SnS. The black arrows for states \uppercase\expandafter{\romannumeral1} and \uppercase\expandafter{\romannumeral3} denote the directions of electric polarization. The green arrows represent for the ferroelastic transition. (b) Energy profile of the kinetic pathway to change the orientations of the electric polarization. (c) Energy spectrum of the SnS nanoflakes under  states \uppercase\expandafter{\romannumeral1}, \uppercase\expandafter{\romannumeral2}, and \uppercase\expandafter{\romannumeral3}. Red dots for \uppercase\expandafter{\romannumeral1} and \uppercase\expandafter{\romannumeral3} represent the in-gap corner states.}
	\label{sns}
\end{figure}

To explicitly testify the existence of fractional corner charges, triangular nanoflakes of BN bilayers are constructed. Figures~\ref{bn}(c) and ~\ref{bn}D display the energy spectrum and distributions of corner states for the AB and BA stackings with sliding ferroelectricity, respectively. The bulk energy gaps are clearly visible and electrons are localized mainly at the $C_3$-symmetric corners, revealing the higher-order topology of ferroelectric BN bilayers. However, interestingly, six corner states emerge and arise from both the up and down layers for the BA stacking, while there are only three corner states that emerge mainly on the up layer for the AB stacking. This is due to the fact that for BA stacking both the up and down layers are terminated with zigzag edges of B atoms, while, for the AB stacking, the up layer is terminated with zigzag edges of B atoms but the down layer is terminated with armchair edges of N atoms. Besides, with the implementation of an out-of-plane electric field, one can achieve a phase transition between AB and BA stacking. Therefore, the exotic corner states can be controlled to be created and annihilated in different layers.

In fact, all the above realizations of ferroelectric HOTIs are the out-of-plane ferroelectric materials with in-plane quantized polarization, we then seek the realizations in in-plane ferroelectric materials. As intrinsic multiferroic materials, $MA$ ($M$ = Ge, Sn; $A$ = S, Se) monolayers have been proposed to be coupled with ferroelasticity and ferroelectricity, enabling the nonvolatile memory readable/writeable capability at ambient condition~\cite{ges_menghao,ges_prl,sns_ex}. As a direct consequence of in-equivalent lattice constants along with the a and b axis, $MA$ have two stable structures, labeled by states \uppercase\expandafter{\romannumeral1} and \uppercase\expandafter{\romannumeral3}, as shown in Fig.~\ref{sns}(a). With the implementation of external in-plane strain or electric field, reversible phase transitions can indeed be obtained between state \uppercase\expandafter{\romannumeral1}-\uppercase\expandafter{\romannumeral3}. By means of the nudged-elastic-band (NEB) method, the overall ferroelastic switching barrier is calculated and presented in Fig.~\ref{sns}(b), almost the same as that of previous theoretical predictions.

To explicitly uncover the HOTI nature, we construct parallelogram nanoflakes of SnS. As shown in Fig.~\ref{sns}(c), two degenerate in-gap states arise around the Fermi level for two ferroelastic states \uppercase\expandafter{\romannumeral1} and \uppercase\expandafter{\romannumeral3}, whereas their accumulation showcase some differences. Effected by polarization, different corners will exhibit different electric potential, rendering a energy splitting between corners. However, the  top-left and down-right corners of state \uppercase\expandafter{\romannumeral1} related by the rotation symmetry $C_{11}$ will share an electric potential of the same amplitude. Therefore, the corner states would emerge as a hallmark of HOTI. Similarly, electrons would accumulate in the top-right and down-left corners of state \uppercase\expandafter{\romannumeral3}, preserving the rotation symmetry $C_{1\bar{1}}$. To be contrast, state \uppercase\expandafter{\romannumeral2} presents us edge states for both the top and bottom edges ([see Fig.~S4~\cite{sm}].) Thus, with the application of strain, one can achieve the transport of corner states, which will give rise to many interseting phenomena and applications.

In conclusion, we have devised a new type of functional phase namely ferroelectric HOTIs and demonstrated that the experimentally synthesized 2D ferroelectrics provide a rich playground to explore. Material candidates are given to traditional ferroelectrics such as In$_2$S$_3$, In$_2$Se$_3$, In$_2$Te$_3$, SnS, GeS, SnSe, GeSe and newly proposed sliding ferroelectrics such as  BN, ZnO, AlN, GaN, SiC, InSe, GaSe~\cite{sm}. We show that the in-plane electric polarizations can give rise to the HOTIs that can be classified into two types with either quantized or non-quantized polarizations. Remarkably, for all of them, electric fields can be used to control the exotic corner states and induce topological phase transitions. Our studies lay the groundwork for electronically controlled higher-order topological phases, and put forward potential material candidates for exploring the intriguing physics.

\vspace{0.3cm}
\noindent{\bf DATA AVAILABILITY}

\noindent
The data that support the ﬁndings of this study are available from the corresponding author upon reasonable request.

\vspace{0.3cm}
\noindent{\bf CODE AVAILABILITY}

\noindent
The codes are available from the corresponding author upon reasonable request.

\vspace{0.3cm}
\noindent{\bf ACKNOWLEDGEMENTS}

\noindent
This work was supported by the National Natural Science Foundation of China (Grants No. 11904205, No. 12074217, and No. 12174220), the Shandong Provincial Natural Science Foundation of China (Grants No. ZR2019QA019 and No. ZR2019MEM013), the Shandong Provincial Key Research and Development Program (Major Scientific and Technological Innovation Project) (Grant No. 2019JZZY010302), and the Qilu Young Scholar Program of Shandong University. 



\vspace{0.3cm}
\noindent{\bf COMPETING INTERESTS}

\noindent
The authors declare no competing interests.

\end{document}